\documentclass[a4paper,12pt] {elsart}
\usepackage[latin1]{inputenc}
\usepackage{graphicx}

\def\L{\mathcal{L}}

\usepackage{amsmath}

\newtheorem{proposicao}{Proposition}

\newtheorem{teorema}{Theorem}

\def\CaixaPreta{\vrule Depth0pt height5pt width5pt}
\def\comecaprova{\noindent {\bf Proof:} \hspace{3mm}}
\def\terminaprova{\hfill \CaixaPreta \vspace{5mm}}
\def\mapright#1#2{\smash{\mathop{\hbox to 0.80cm{\rightarrowfill}}\limits^{#1}_{#2}}}
\def\mapleft#1#2{\smash{\mathop{\hbox to 0.80cm{\leftarrowfill}}\limits^{#1}_{#2}}}
\def\mapleftright#1#2{\smash{\mathop{\hbox to 0.80cm{\leftarrowfill \rightarrowfill}}\limits^{#1}_{#2}}}

 \def\red#1{\lfloor #1 \rfloor}
 \def\se#1{#1 \hspace{-3.82mm}{\raise 1.8mm  \hbox{ {\tiny $\rightharpoonup$} } }\hspace{-1.6mm} }
 \def\Sorted#1{#1 \hspace{-3.82mm}{\raise 2.8mm  \hbox{ {\tiny $\rightharpoonup$} } }\hspace{-1.7mm} }

 \def\ceila#1{ \lceil #1 \rceil^a }
 \def\w#1{\hspace{-1.5mm} \raise 2.2mm  \hbox{ {\tiny $\leftarrow$} } \hspace{-4.0mm} #1 }
 \def\sw#1{\hspace{-1.5mm} \raise 1.0mm  \hbox{ {\tiny $\swarrow$} } \hspace{-2.0mm} #1 }
 \def\o#1{\overline{#1}}
 \def\nat{I\hspace{-1.3mm}N}
 \def\se#1{\hspace{0 mm} #1 \hspace{-2.8mm} \raise 1.6mm  \hbox{ {\tiny $\searrow$} \hspace{-2.0mm}} }
 \def\Se#1{\hspace{0 mm} #1 \hspace{-2.8mm} \raise 2.5mm  \hbox{ {\tiny $\searrow$} \hspace{-1.5mm}} }

\begin{document}

\begin{frontmatter}

\title{PHORMA: Perfectly Hashable Order Restricted Multidimensional Arrays}
\author[A1]{Lauro Lins}
\author[A2]{\hspace{-1mm}, Sóstenes Lins}
\author[A2]{\hspace{-1mm}, Sílvio Melo}
\address[A1]{Centro de Informática - UFPE - Recife - Brazil}
\address[A2]{Departamento de Matemática - UFPE -
Recife - Brazil}

\bibliographystyle{alpha}
\maketitle

\begin{abstract}
In this paper we propose a simple and efficient data structure
yielding a perfect hashing of quite general arrays. The data
structure is named {\em phorma}, which is an acronym for {\em
p}erfectly {\em h}ashable {\em o}rder {\em r}estricted {\em
m}ultidimensional {\em a}rray.

\begin{description}
\item [Keywords:] {Perfect hash function, Digraph, Implicit
enumeration, Nijenhuis-Wilf combinatorial family.}

\item [AMS-class:] 05A05; 05C90; 06F99. \item [ACM-class:] E2; E1.

\end{description}

\end{abstract}
\end{frontmatter}

\section{Motivation}
\label{sec:Motivation}

\small

Let $a=a_1a_2\ldots a_n$ and $\alpha=\alpha_1\alpha_2\ldots a_n$
be $n$-sequences of positive integers, $\alpha\le a$, meaning
$\alpha_ i \le a_i, i=1,2,\ldots,n.$ Suppose that $f(\alpha)$ is a
{\em symmetric function} on the variables $\alpha_i$, that is, the
value of $f(\alpha)$ does not change if the coordinates of
$\alpha$ are permuted in an arbitrary way. To store the function
$f$, it is enough to allocate space for the values of $f(\alpha)$,
where $\alpha_i \ge \alpha_{i+1}$, $1 \le i \le n-1$. Thus, we
need to enumerate the $\alpha$'s satisfying $\alpha \le a$ and the
boolean function
$$B^{n\ge}_{sym}= (\alpha_1 \ge \alpha_2) \wedge (\alpha_2 \ge
\alpha_3) \wedge \ldots \wedge \ldots (\alpha_{n-1} \ge
\alpha_n).$$

The motivation for this work is to enumerate and give a perfect
hash function \cite{CormenLeisersonRivest1990,Knuth1975} for
multidimensional arrays which have order restrictions on their
entries. The simplest example of this situation is when the
restrictions are given by $B_{sym}^{n\ge}$. We show that quite
general boolean functions can take the place of $B_{sym}^{n\ge}$
and that the large class of enumerative/perfect hash associated
problems can be put under a common framework.

    \begin{figure}
        \begin{center}
            \includegraphics{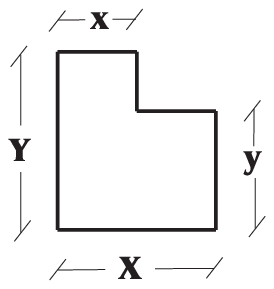}
            \caption{\sf The $L$-piece}
            \label{fig:Lpiece}
        \end{center}
    \end{figure}

To exemplify the appearance of a more complex boolean function,
consider the problem of efficiently enumerate all the $L$-shaped
pieces with vertices which fit in a $(p \times q)$ integer grid.
This is a typical situation treated in
\cite{LinsLinsMorabito2002B}. An {\em $L$-shaped piece} is a
rectangle $R$ from which we have removed a smaller rectangle $r
\subseteq R$. Moreover $R$ and $r$ have a corner in common. By
effecting rotations, translations and reflections we may suppose
that our $L$-shaped piece has a corner in the origin and the
common vertex to $r$ and $R$ is the vertex opposite to the origin
in rectangle $R$. Positioned in this way, the $L$-piece is
represented by a quadruple of positive integers
$(X,Y,x,y)=\alpha_1\alpha_2 \alpha_3 \alpha_4\le
a_1a_2a_3a_4=(p,q,p,q)$, as in Figure \ref{fig:Lpiece}.

The geometry imposes the restrictions: (1) $X \ge x$; (2) $Y \ge
y$. Symmetry considerations enable us to partition the set of
$a$-bounded $L$-pieces into equivalent classes and to distinguish
a set $A$ of representatives for these classes. For the occupancy
purposes in \cite{LinsLinsMorabito2002B}, the $L$-pieces
$(X,Y,x,y)$ and $(Y,X,y,x)$ must be considered equivalent. This
implies the restrictions: (3) $X \ge Y$ and (4) $X=Y \Rightarrow x
\ge y$. In terms of occupancy, $(X,Y,X,y)$ with $y<X$, which is a
degenerated $L$, can (and must) be replaced by the rectangle
$(X,Y,X,Y)$. Analogously, $(X,Y,x,Y)$ with $x < X$ must be
replaced by $(X,Y,X,Y)$. In this way, the equivalence $(X=x)
\Leftrightarrow (Y=y)$ holds. The equivalence is rewritten as two
opposite implications in the disguised form: (5) $(X \ne x) \ \vee
\ (Y=y)$ and (6) $(Y \ne y) \ \vee \ (X=x)$. The restrictions (1)
to (6) are gathered in a boolean expression $B_L$ in terms of the
$\alpha_i$'s:

\begin{center}
$B_L = (\alpha_1 \ge \alpha_3) \wedge (\alpha_2 \ge \alpha_4)\
\wedge \ (\alpha_1 \ge \alpha_2)\ \wedge \ ((\alpha_1 \ne
\alpha_2)\ \vee\
(\alpha_3 \ge \alpha_4)) \ \wedge \ $ \\
$ ((\alpha_1 \ne \alpha_3)\  \vee \ (\alpha_2=\alpha_4))\ \wedge \
((\alpha_2 \ne \alpha_4) \ \vee \ (\alpha_1=\alpha_3)).$
\end{center}

So, we want to enumerate the $4$-sequences
$\alpha=\alpha_1\alpha_2\alpha_3\alpha_4$ of positive integers
$\alpha \le a$ and satisfying $B_L$. If, as it is typically needed
in packing problems, $a$ is of order
$(120,100,120,100)=(120,100)^2$ then we have $23,094,225$
$\alpha$'s that satisfies $B_L$ in a total of $144,000,000$
possibilities. If $a=(7,5)^2$, then there is a total of $190$
$\alpha$'s in $1225$ possibilities. The valid $190$ $\alpha$'s are
in $1-1$ correspondence with the $st$-paths in the digraph of
Figure \ref{fig:G7575BL}.

\section{The Definition of Phorma and the Objective of the Work}
\label{sec:DefPhormaObjWork}

Let $\nat$ be the set of natural numbers, $\nat^\star=\nat
\backslash \{0\}$ and $N=\{1,2,\ldots,n\}$. For $1 \le m \le n$,
define $M=\{1,\ldots,m\}$. Let $Y^X$ be the set of all functions
from $X$ into $Y$. Throughout this work,
$\alpha=\alpha_1\ldots\alpha_n$ is an $n$-sequence of positive
integers, that is, $\alpha \in (\nat^\star)^N$. The relation
$\rho' \le \rho$ for sequences $\rho'$ and $\rho$ of equal length
means that $\rho'_i \le \rho_i$, for each $i$-term of the
sequences.

An {\em $n$-composition} $\delta=\delta_1\ldots\delta_m$ is an
element of $(\nat^\star)^M$ such that $\sum_{1\le m \le
n}\delta_m=n$. The set of $n$-compositions is denoted by $C^n$.
Given $\alpha$, let $m_\alpha$ be the number of distinct entries
in $\alpha$ and $m_\delta$ be the length of $\delta$. Let
$\o{\alpha}=\o{\alpha}_1\ldots\o{\alpha}_{m_\alpha} \in C^n$
denote the $n$-composition where $\o{\alpha}_i$ is the number of
occurrences of the $i$-th smallest entry of $\alpha$.

An {\em $n$-phorma} is a triple $P=(a,B,C)$ satisfying: (i)
$a=a_1a_2\ldots a_n \in (\nat^\star)^N$; (ii) $B$ is a boolean
function whose literals of $B$ are of type $(\alpha_i \star
\alpha_j)$, where $\alpha \in (\nat^\star)^N$ and $\star \in \{
\le, \ge, <,
>, =, \ne\}$; (iii) $C \subseteq C^n$ is a given set of
$n$-compositions. The term $n$-phorma is an acronym for an {\em
$n$}-dimensional {\em p}erfectly {\em h}ashable {\em o}rder {\em
r}estricted {\em m}ultidimensio\-nal {\em a}rray.

The objective of this paper is to enumerate the set
$$A(P)=A(a,B,C) = \{ \alpha \ | \ \alpha \le a,
\ \alpha {\rm \ satisfies\ } B,\ \o{\alpha} \in C\}.$$ In the
particular case that $B$ is the empty boolean function, then there
are no $B$-restrictions and $A(a,B,C)$ is the subset of
$(\nat^\star)^N$ consisting of all sequences $\alpha \le a,\
\o{\alpha} \in C$. We construct a bijection $ h: A(P)
\longrightarrow \{0,1,\ldots,|A(P)|-1\},$ so that both $h$ and
$h^{-1}$ are efficiently computable. Such functions are called
{\em perfect hash functions}
\cite{CormenLeisersonRivest1990,Knuth1975}. Their usefulness is
well known.

As far as we know the problem of finding perfect hash functions
for these quite general multidimensional arrays have not been
considered before in the literature, whence the lack of more
specific references and bibliography. Our solution is based on the
theory of {\em combinatorial families} developed in
\cite{NijenhuisWilf1978}. Here we call these families {\em
NW-families} and recall their definition in Section
\ref{sec:NWFamilies}. The central idea is to associate a digraph
to a collection of combinatorial objects in such a way that each
object in the family is in $1-1$ correspondence with a path in the
digraph. A more detailed account of these combinatorial families
appears in \cite{Wilf1990}.

From a phorma $(a,B,C)$ a digraph $G(a,B,C)$ with a single source
$s$ and a single sink $t$ can be constructed so that the elements
in $A(a,B,C)$ are in $1-1$ correspondence with the $st$-paths.
Indeed, $G(a,B,C)$ is {\em an NW-family} \cite{NijenhuisWilf1978}
encoding $A(a,B,C)$ with a simple perfect hash function $h$. We
briefly review these families in Section \ref{sec:NWFamilies}. The
digraph $G((7,5)^2,B_L,C^4)$ associated to the phorma
$((7,5)^2,B_L,C^4)$ is shown in Figure \ref{fig:G7575BL}. In this
example, the set $C$ of $4$-compositions is the whole set $C^4$.

\section{More Applications of Phormas}

\begin{figure}
        \begin{center}
            \includegraphics{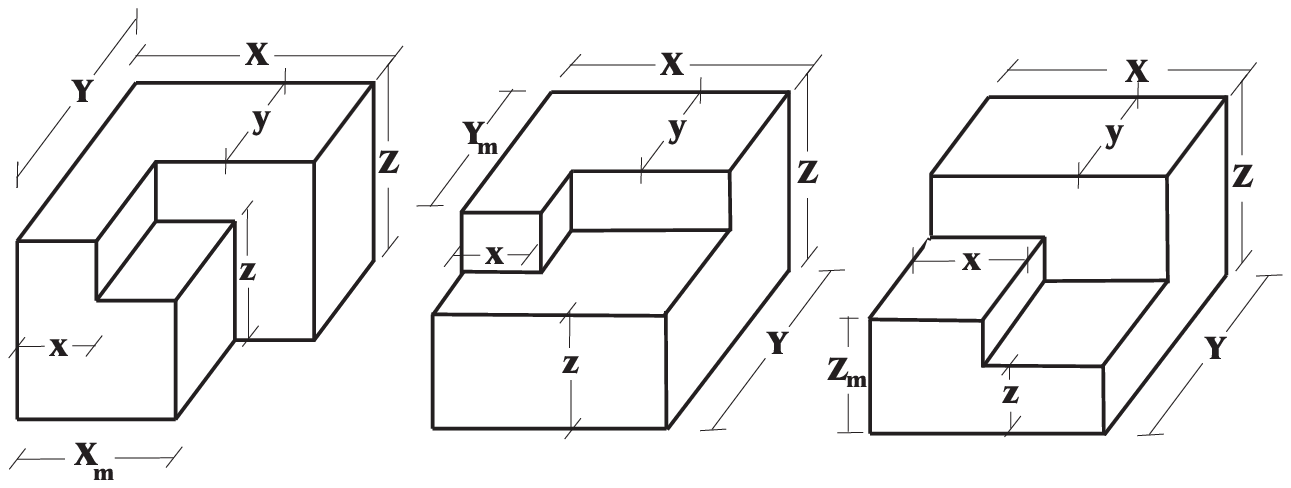}
            \caption{\sf The $T$-pieces $T_x$, $T_y$ and $T_z$}
            \label{fig:Tpiece}
        \end{center}
    \end{figure}

The need to impose order restrictions on arrays appears frequently
and in many cases it is not difficult to express these
restrictions as a phorma. For a larger example, consider the
$7$-phormas arising from the generation of {\em $T$-shaped
pieces}. In Figure \ref{fig:Tpiece} we show the three kinds of
such a piece. They are composed of a $3$-block and a $3D$
$L$-piece. In the case of the $T_z$-piece, the $L$ is truncated in
one of its legs along the $z$-direction. These pieces are the $3D$
counterpart for the $2D$ $L$-shaped piece and they play an
important role in $3D$ packing problems. They are described by
seven parameters, which in the case of the $T_z$-piece are,
$(x,X,y,Y,z,Z_{m},Z)$. To enumerate the $T$-pieces contained in a
$(p \times q \times r)$-block was the motivating idea to formalize
the notion of phorma. The need to effect this enumeration appears
in \cite{Lins2003}.

As an example , for the $T_z$-piece, the restrictions coming from
the geometry and the symmetry on the seven parameters
$xXyYzZ_mZ=\alpha_1\alpha_2\alpha_3\alpha_4\alpha_5\alpha_6\alpha_7$
are of three types:
 \begin{enumerate}
    \item $(X \ge x);  (Y \ge y); (Z \ge Z_m \ge z)$;
    \item $(X \ge Y); (X=Y) \Rightarrow (x \ge y)$;
    \item $(x = X) \Rightarrow (z = Z_m);  (y=Y) \Rightarrow (x=X) \wedge (z=Z)$.
 \end{enumerate}

    \vspace{0.5cm}

The first type of restrictions is obvious. The second type
expresses the fact that the $T_z$-piece can be rotated around a
vertical axis without modifying its containment properties. The
$X$- and $Y$-directions are equivalent. Other axis of rotations,
implying similar restrictions, could be used if the boxes to be
packed into the $T_z$-piece could change its vertical. The third
type of restrictions deals with the degenerated cases, in which
the $T_z$-piece becomes a simpler piece. In terms of a phorma type
boolean function, the restrictions translate as a boolean function
$B_T^z$ with the following $9$ clauses:

   $$B^z_T=(\alpha_2 \ge \alpha_1)\wedge (\alpha_4 \ge \alpha_3)\wedge (\alpha_7 \ge \alpha_6)
   \wedge (\alpha_6 \ge \alpha_5)\wedge
   (\alpha_2 \ge \alpha_4)\wedge$$
   $$((\alpha_2 \neq \alpha_4) \vee (\alpha_1 \ge
   \alpha_3))\wedge ((\alpha_1 \neq \alpha_2) \vee (\alpha_5 = \alpha_6))\wedge$$
   $$ ((\alpha_3 \neq \alpha_4) \vee (\alpha_1 = \alpha_2))\wedge
    ((\alpha_3 \neq \alpha_4) \vee (\alpha_5 = \alpha_7)).$$

In this case, once more, $C$ is the whole set of $7$-compositions
$C^7$. If, just to be specific, $a=(15^217^219^3)$, then
$|A(a,B_T^z,C^7)|=7,510,130$, while $15^2 17^2 19^3=446,006,475.$
The amount of memory required to store the digraph
$G(a,B_T^z,C^7)$ is logarithmically smaller than
$|A(a,B_T^z,C^7)|$ (see Figure \ref{fig:GParameters}) and its
construction takes only a few seconds of computer time. Along the
same line, we can derive boolean functions $B_T^x$ and $B_T^y$ for
the other $T$-pieces $T_x$ and $T_y$ shown in Figure
\ref{fig:Tpiece}. The three $T$-pieces are inequivalent under
reflections and rotations which maintain the vertical direction.
They play a complementary role in $3D$ packing problems in which
the vertical direction of the boxes to be packed must be
preserved.

We briefly mention another application of phorma: finding all the
solutions for {\em Cube It}. Let $x<y<z$ be real numbers. Consider
the problem of finding all maximum packings of $(x \times y \times
z)$-bricks into a cube of side $x+y+z$. If $y+z < 3x$, then $27$
is an upper bound on the number of bricks that can be packed, see
\cite{Hoffman1981}. There exists a phorma
$(3^{81},B^{Cube}_{It},c^3_{27})$ of dimension $81$ such that
$A(3^{81},B^{Cube}_{It},c^3_{27})$ has 1008 elements coinciding
with the 1008 distinct solutions for the problem of packing the
maximum of $27$ boxes. In this case, $a$ is the sequence of $81$
repetitions of $3$, $a=3333\ldots3$ and $C=\{c^3_{27}\}$, where
$c^3_{27}=(27,27,27)$. The expression for $B^{Cube}_{It}$ and its
justification are too long to be included in this paper. A higher
dimensional analogue of $B^{Cube}_{It}$ relates to an interesting
open problem which is the subject of ongoing research: how to pack
$5^5=3125$ $(a\times b \times c \times d \times e)$-boxes into a
$5$-cube of side $a+b+c+d+e$. Our implementation (not yet
optimized) of the phorma $(3^{81},B^{Cube}_{It},c^3_{27})$ found
the 1008 solutions in about a day of computer time. What is
interesting to mention, is that there are no symmetries in these
1008 solutions. So, their set  can be partitioned into $21$
classes of $48$ elements each, corresponding to the symmetry group
of the cube. Representatives of these $21$ classes are given in
Figure \ref{fig:solutionForCubeIt}. The bricks orientations are
given by the conventions: $ a \mapsto yzx$; $ A \mapsto zyx$; $  b
\mapsto xzy$; $ B \mapsto zxy$; $ c \mapsto xyz$; $ C \mapsto
yxz.$ The parameters of $G(3^{81},B^{Cube}_{It},c^3_{27})$ are
listed in Figure \ref{fig:GParameters}. In particular,
$H(3^{81},B^{Cube}_{It},c^3_{27})$ has only $4$ vertices and the
whole difficulty is to find
$\red{A(3^{81},B^{Cube}_{It},c^3_{27})}$ which in this case
coincides with $A(3^{81},B^{Cube}_{It},c^3_{27})$.

\begin{figure}
\begin{center}
{\scriptsize
\begin{tabular}{|c|c|c|c|c|c|c|c|c|c|c|} \hline
sol 1 & sol 2 & sol 3 & sol 4 & sol 5 & sol 6 & sol 7 & sol 8 &
sol 9 & sol 10 & sol 11 \\ \hline \hline
acC & acC & acC & acC & acC & acC & acC & acC & acC & acB & acB \\
bCA & bBc & bBc & bBc & bBc & bBA & bAB & bAB & bAB & bCc & bCc \\
ABb & AaB & AaB & AaB & AaB & Bac & Bac & Bac & Bac & BAa & BAa \\
\hline

cbB & bBA & bBA & bAB & caB & cbB & cbB & cbB & cbB & caC & cbC \\
aAC & ACa & ACb & ACb & CAb & aAC & aCc & aCc & aBc & AbB & aBA \\
Bac & Cbc & Cca & Cca & Bca & ACb & ABa & ABa & ACa & aBc & ACb \\
\hline

BaA & caB & caB & cBa & bBA & BaA & BaA & BAa & BaA & BbA & BAa \\
cBb & CAb & CAa & CaA & ACa & Ccb & cBa & CbA & Cca & CAa & caB \\
Cca & Bca & Bbc & Bbc & Cbc & cBa & Ccb & cCb & cBb & cCb & Cbc \\
\hline
\end{tabular} }
\end{center}

\begin{center}
{\scriptsize
\begin{tabular}{|c|c|c|c|c|c|c|c|c|c|} \hline
sol 12 & sol 13 & sol 14 & sol 15 & sol 16 & sol 17 & sol 18 & sol
19 & sol 20 & sol 21 \\ \hline \hline
acB & acB & acB & acB & acB & acB & acB & acB & aBc & aBc  \\
bCc & bCc & bCc & bCc & bCc & bCc & bCc & bAC & AaC & AaB  \\
BAa & BAa & BAa & ABa & ABa & ABa & ABa & Bac & cbB & bcC  \\
\hline

cbC & cbC & cbC & bAC & bAC & cbC & cbC & cbC & BAa & CAa  \\
aAB & AaB & AaB & AaB & AbB & aBA & aAB & aBA & bcB & Bbc  \\
ACb & aBc & aBc & Ccb & Cac & Bac & Bac & ACb & aCc & caB  \\
\hline

BaA & BaA & BAa & cBa & cBa & BAa & BaA & BaA & Ccb & Bcb \\
cBa & cBa & CbA & CbA & CaA & caB & cBa & Ccb & cBa & cCa \\
Cbc & Ccb & cCb & Bac & Bcb & Ccb & Ccb & cBa & BaA & aBA \\
\hline
\end{tabular} }
            \caption{\sf All the solutions for Cube It}
            \label{fig:solutionForCubeIt}
\end{center}
\end{figure}

\section{NW-Families}
\label{sec:NWFamilies}

The following concept, introduced in \cite{NijenhuisWilf1978}, is
the central tool for our hashing scheme. A {\em Nijenhuis-Wilf
combinatorial family}, or simply an {\em NW-family}, is an acyclic
digraph $G$ whose vertex set is denoted by $V(G)$, having the
properties below:
\begin{enumerate}
\item $V(G)$ has a partial order (for $x,y \in V(G)$, $y \preceq
x$ if there is a directed path from $x$ to $y$) with a unique
minimal element $t$. For each $v \in V(G)$ the set $\{x \in V(G) \
| \ x\preceq v\}$ is finite and includes $t$. \item Every vertex
$v$, except $t$ has a strictly positive outvalence $\rho(v)$. For
each $v \in V(G)$, the set $E(v)$ of outgoing edges has a $v${\em
-local rank-label} $\ell_v$, $0 \le \ell_v(e)\le \rho(v)-1, e \in
E(v)$.
\end{enumerate}

A path starting at $v$ and ending in $t$ {\em is encoded} by the
sequence of label-ranks of the sequence of its edges. Such a path
is called an object of {\em order} $v$ \cite{NijenhuisWilf1978}.
The beauty of this scheme is that we can perform various tasks on
the family in an abstract way, without referring to the actual
encoding/decoding of the objects as paths. An NW-family is
especially suited to deal with the following $5$ tasks. Tasks $1$
to $4$ are from \cite{NijenhuisWilf1978}. Task $0$ is emphasized
here because of its applicability to the phorma: we need to
calibrate the cardinality of $A(a,B,C)$ by choosing $a$ in an
adequate way.

\noindent {\sf Task 0: counting}: What is the family's cardinality?
 {\sf Algorithm:} Given $v \in V(G)$, let $|v| = \sum
\{|head(e)| \ | \ e \in E(G), tail(e)=v\}.$ From this formula,
$|v|$ is easily obtained by recursion. It is convenient to store
it as an attribute of $v \in V(G)$ in a pre-processing phase, or
{\em compilation time.}

\noindent {\sf Task 1: sequencing}: Given an object in the family,
construct the ``next'' object. {\sf Algorithm:}  The {\em next
path} of a given path $\pi$ in coded form is, in coded form, the
lexicographic successor of $\pi$.

\noindent {\sf Task 2: ranking (perfect hashing)}: Given an object
$\omega$ in the family, find the integer $h(\omega)$ such that
$\omega$ is the $h(\omega)$-th element in the order induced by
task 1. {\sf Algorithm:} Let an element-path $\pi$ of order $v$ of
an NW-family, $\pi=(e_1,e_2,\ldots,e_p)$ be given. The rank of
$\pi$ is defined as $h(\pi) = \sum_{i=1}^p \chi(e_i)$, where
$\chi(e)=\sum \{|head(f)| \mbox{ with } \ell_v(f) < \ell_v(e),f
\in E(v) \}$.

\noindent{\sf Task 3: unranking}: Given an integer $r$, we need to
construct the $r$-th path from $v$ to $t$. Define $pred_v(e)$ as
the highest-rank edge of the set $\{f\in E(v)\ | \ \ell_v(f) <
\ell_v(e) \}$, and let $|head(pred_{v}(e))|=0$ if this set is
empty. The required $r$-th path $\pi_r$ is generated as follows.
{\sf Algorithm:} $\pi_r \leftarrow \emptyset$; $r' \leftarrow 0$;
$v' \leftarrow v$; {\bf repeat} append to $\pi_r$ the highest-rank
edge $e$ of $E(v')$ such that $r'+|head(pred_{v'}(e))| \le r$; $r'
\leftarrow r'+|head(pred_{v'}(e))|$; $v' \leftarrow head(e)$ {\bf
until} $v'=t$.

\noindent {\sf Task 4: getting random object}: Choose an object
uniformly at random from the given family. {\sf Algorithm:} Let
$\xi \in [0,1]$ be uniformly chosen at random; return the
$(|v|*\xi)$-th object.

\section{Reducing, Sorting, $a$-Roofing: the Digraph $G(a,B,C)$}
\label{sec:RedSortRoofDigraph}

If $\alpha$ has $m \le n$ distinct entries, let $M_\alpha
=\{1,\ldots, m\}$. The {\em reduction} of $\alpha$, denoted by
$\red{\alpha}$, is the unique surjection in $(M_\alpha)^N$ which
is order compatible with $\alpha$. That is, for $i \in N$, if
$\alpha_i$ is the \hbox {$j$-th} smallest entry in $\alpha$, then
$\red{\alpha}_i=j$. Let also $\se{\alpha}$ denote the $m$-sequence
of distinct entries of $\alpha$ in ascending order. We call
$\se{\alpha}$ the {\em sorting} of $\alpha$. Given an ascending
$m$-sequence $\gamma$, let $m_\gamma=m$.

\begin{proposicao} The $n$-vector of positive integers $\alpha$ is recoverable from
$(\red{\alpha},\se{\alpha})$. \label{proposicao:recoverability}
\end{proposicao}

\comecaprova It is sufficient to observe that  $\alpha_i =
\se{\alpha}_{\red{\alpha}_i}$. \terminaprova

Since $\alpha$ induces the pair $(\red{\alpha},\se{\alpha})$ and,
by Proposition \ref{proposicao:recoverability}, is recoverable
from, it we can think of $\alpha$ as the pair
$(\red{\alpha},\se{\alpha})$ and write $\alpha \equiv
(\red{\alpha},\se{\alpha})$.

For $\alpha \in A(a,B,C)$ let the {\em $a$-roof} of $\alpha$ be
$\lceil \alpha \rceil ^a = \gamma^\star$ where $\gamma^\star$ is
the lexicographically maximal increasing $m$-sequence with the
property that $(\red{\alpha},\gamma^\star) \in A(a,B,C)$. In
particular, $\se{\alpha}_i \le \ceila{\alpha}_i=\gamma^\star_i$,
$i \in N$.

\begin{proposicao} The $a$-roof of $\alpha$, $\ceila{\alpha}$, does not depend
on $\alpha$ itself but only on $\red{\alpha}$ and $a$, in the
sense that $\ceila{\alpha} = \ceila{\red{\alpha}}$.
\label{proposition:classdependency}
\end{proposicao}

\comecaprova The $a$-roof $\ceila{\alpha} =
\gamma_1^\star\gamma_2^\star\ldots\gamma_m^\star$ can be
constructed as follows. Suppose that, for $1\le i \le m$, $i$
occurs at positions $p_{i1}, \ldots, p_{ij_i}$ of $\red{\alpha}$.
Then we must have $\gamma_m^\star=\min \{ a_{p_{m1}}$,
$a_{p_{m2}}$, $\ldots$, $a_{p_{mj_m}}\}$, due to $a$-dominance.
For $i=m-1, m-2, \ldots,1$, the definition implies that
$\gamma_i^\star=\min \{$ $a_{p_{i1}}, \ldots,
a_{p_{ij_i}},\gamma_{i+1}^\star-1 \}$, by $a$-dominance and to
insure the strict increase of $\gamma^\star$. Since the
construction only depended on $\red{\alpha}$ and $a$, the
Proposition is proved. \terminaprova

Given a phorma $(a,B,C)$ and the corresponding $A(a,B,C)$, three
sets are defined:
   $(i)\ \red{A(a,B,C)} = \{ \red{\alpha} \ | \ \alpha \in A(a,B,C)\},$
   $(ii)\ \Se{A}(a,B,C) = \{ \se{\alpha} \ | \ \alpha \in A(a,B,C)\},$
   $(iii)\ \lceil{A(a,B,C)}\rceil^a = \{ \ceila{\alpha} \ | \ \alpha \in A(a,B,C)\}.$

Usually, but not necessarily (see the phorma
$(3^{81},B^{Cube}_{it},c_{27}^3)$), $|\red{A(a,B,C)}|$ is much
smaller than $|A(a,B,C)|$. By Proposition
\ref{proposition:classdependency}, $|\lceil{A(a,B,C)}\rceil^a| \le
|\red{A(a,B,C)}|.$ In general this inequality is also not tight.
See examples in Figure \ref{fig:GParameters}. The perfect hash
function that is constructed for $A(a,B,C)$ depends on the
explicit enumeration of the set $\red{A(a,B,C)}$. This set, in the
case of our ongoing example, has nine elements,
$$\red{A((7,5)^2,B_L,C^4)} =
\{1111,2121,2211,3211,3221,3321,4231,4312,4321\}.$$ The $a$-roof
set has only seven elements because of two duplicates
$$\lceil A((7,5)^2,B_L,C^4)\rceil^a =
\{5,57,45,457,457,345,4567,3457,3457 \}.$$

\begin{figure}[!h]
        \begin{center}
            \includegraphics{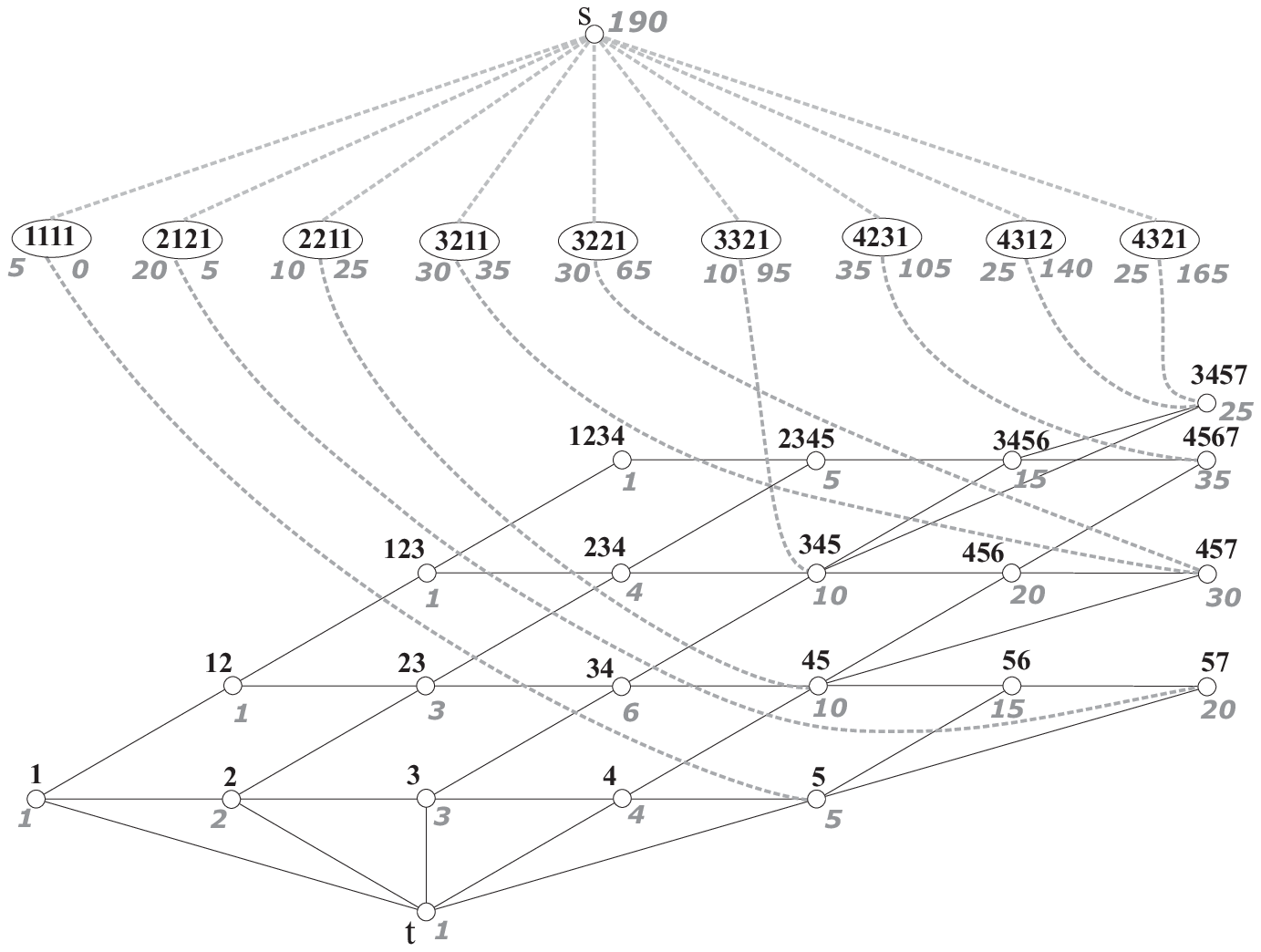}
            \caption{{\sf Digraph $G(7575,B_L,C^4)$ encoding
                      $A(7575,B_L,C^4)$}}
            \label{fig:G7575BL}
        \end{center}
    \end{figure}

Given a phorma $(a,B,C)$ the {\em digraph $\Lambda(a,B,C)$} is
defined as follows. Its vertex set is $V(\Lambda(a,B,C))= \{s\}
\cup \red{A(a,B,C)} \cup \lceil{A(a,B,C)}\rceil^a$, where $s$ is a
single source. It is a simple graph, and so, each of its directed
edges can be represented by an ordered pair of vertices. For each
$\red{\alpha} \in \red{A(a,B,C)}$ there are an edge $(s,
\red{\alpha})$ and an edge $(\red{\alpha}, \ceila{\alpha})$. These
are all the edges of $\Lambda(a,B,C)$, concluding its definition.
The digraph $\Lambda(a,B,C)$ is a subgraph of $G(a,B,C)$. In
Figure \ref{fig:G7575BL}, the edges of $\Lambda((7,5)^2,B_L,C^4)$
are depicted in dashed gray. The edges of its complement
$H((7,5)^2,B_L,C^4)$ in  $G((7,5)^2,B_L,C^4)$ (which we define
next) are depicted in solid lines.  The number near a vertex $v$
(the first number, when there are two) is the number of $vt$-paths
in $G((7,5)^2,B_L,C^4)$.

Let $H^\infty$ be the set of all finite strictly increasing
sequences of positive integers. The empty sequence is in
$H^\infty$ and is denoted by $t$. Suppose
$\gamma=\gamma_1\gamma_2\ldots\gamma_m \in H^\infty$. We define an
NW-family $H_\gamma$ as follows. If $\gamma_m > m$ let
$\w{\gamma}$ denote the increasing sequence of length $m$
satisfying $\w{\gamma}_m=\gamma_m-1$ and $\w{\gamma}_i=\min \{
\w{\gamma}_{i+1}-1,\gamma_i\}$, for $i=m,m-1,\ldots,1$. If
$\gamma_m=m$, then $\w{\gamma}$ does not exist. If $\gamma\ne t$,
let $\sw{\gamma}$ be the sequence of length $m-1$ obtained from
$\gamma$ by removing its last entry:
$\sw{\gamma}=\gamma_1\ldots\gamma_{m-1}$. If $\gamma = t$, then
$\sw{\gamma}$ does not exist. Given $\overline{\gamma},
\widetilde{\gamma} \in H^\infty$, we say that $\overline{\gamma}
\preceq \widetilde{\gamma}$, if there is a sequence
$(\widetilde{\gamma}=\gamma^1, \gamma^2, \ldots,
\gamma^p=\overline{\gamma})$, with $\gamma^i \in H^\infty$, such
that, for each $i=1,2,\ldots,p-1$, either
$\gamma^{i+1}=\w{\gamma}^{i}$ or else
$\gamma^{i+1}=\sw{\gamma}^{i}$. The relation $\preceq$ makes
$H^\infty$ a partial ordered set, or {\em poset}. For $\gamma \in
H^\infty$, let $H_\gamma$ be the acyclic digraph whose vertex set
is $V(H_\gamma) = \{ \gamma' \ | \ \gamma' \preceq \gamma\}$. From
each vertex $\gamma' \in V(H_\gamma)$ there are at most two
outgoing edges: $(\gamma', \w{\gamma}')$, of $\gamma'$-local
rank-label $0$, if $\w{\gamma}'$ exists and $(\gamma',
\sw{\gamma}')$, if $\sw{\gamma}'$ exists. The $\gamma'$-local
rank-label of this last edge is either $1$, if $\w{\gamma}'$
exists or $0$ otherwise. This concludes the definition of
$H_\gamma$.

Given a path $\pi$ from $\gamma^\star$ to $t$ in
$H_{\gamma^\star}$, a {\em fall} of $\pi$ is a vertex $\gamma$
such that the edge $(\gamma,\sw{\gamma})$ is used by $\pi$. Path
$\pi$ has exactly $m_{\gamma^\star}$ falls. In Figure
\ref{fig:H5679} the $4$ falls of the path shown in thick edges
are: $5678$, $567$, $34$ and $3$. The encoding/decoding of the
increasing sequences $\gamma \preceq \gamma^\star$ as paths in the
NW-family $H_{\gamma^\star}$ is particularly simple:

    \vspace{0.5cm}
    \begin{figure}[!h]
        \begin{center}
            \includegraphics{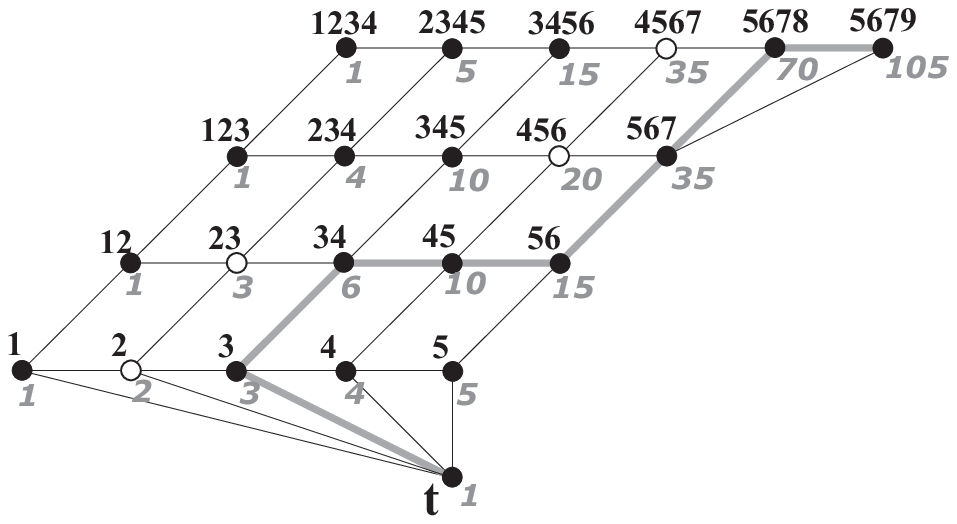}
            \caption{{\sf Path $\pi$ in $H_{5679}$ with falls $5678$, $567$, $34$ and $3$ encoding $\gamma=3478$}, $h(\gamma)=60$}
            \label{fig:H5679}
        \end{center}
    \end{figure}

\begin{proposicao} To a path $\pi$ in $H_{\gamma^\star}$ from
$\gamma^\star$ to $t$ corresponds $\gamma_\pi \preceq
\gamma^\star$ consisting of the last coordinates of the $\pi$
falls (in reverse order). Reciprocally, to  $\gamma \preceq
\gamma^\star$, corresponds the unique path $\pi_\gamma$ from
$\gamma^\star$ to $t$ such that the last entry of its $i$-th fall
coincides with the $i$-th entry of $\gamma$. Moreover,
$\pi_{\gamma_\pi}=\pi$, $\gamma_{\pi_\gamma}=\gamma$.
\label{proposition:EncodingDecoding}
\end{proposicao}
\comecaprova Straightforward from the definitions. \terminaprova

Given a path $\pi$ from $\gamma^\star$ to $t$ in
$H_{\gamma^\star}$, a {\em post-fall} of $\pi$ is a vertex
$\gamma'=\w{\gamma}$ such that the edge $(\gamma,\sw{\gamma})$ is
used by $\pi$. Path $\pi$ has at most $m_{\gamma^\star}$ falls.
The set of post-falls of $\pi$ is denoted $PostFall(\pi)$. In
Figure \ref{fig:H5679}, calling $\pi$ the path shown in thick
edges, we have $PostFall(\pi)= \{4567, 456, 23,2\}$ and their
members are depicted as white vertices. The hash function
$h_{\gamma^\star}$ in the NW-family $H_{\gamma^\star}$ takes a
simple form:

\begin{proposicao} The perfect hash function associated with the NW-family $H_{\gamma^\star}$
is $h_{\gamma^\star}(\gamma) = \sum\{ |\gamma'| \ \ | \ \gamma'
\in PostFall(\pi_\gamma) \}.$
 \label{proposition:RankingUnranking}
\end{proposicao}
\comecaprova The result is an specialization of the rank function
of a generic NW-family to $H_{\gamma^\star}.$ It follows directly
from the definitions. \terminaprova

From this Proposition it follows in Figure \ref{fig:H5679} that
$h_{5679}(3478)=35+20+3+2=60$. The terms of the sum correspond to
the orders of the white vertices, forming the set
$PostFall(\pi_{3478})$.

Define $H(a,B,C) = \bigcup \{H_{\gamma^\star} \ | \ \gamma^\star
\in \lceil{A(a,B,C)}\rceil^a \}.$ Actually, in this union we need
only to take maximal $\gamma^\star$'s. If $\gamma' \preceq
\gamma^\star$, then $H_{\gamma'}$ is a subgraph of
$H_{\gamma^\star}$ and it is irrelevant for the union. The digraph
$H(7575,B_L,C^4)$ shown in Figure  \ref{fig:G7575BL}, is formed by
the union of 4 maximal $\gamma^\star$'s: $H_{3457}\cup H_{4567}
\cup H_{457} \cup H_{57}$. In general, the {\em digraph of a
phorma} $P=(a,B,C)$ is defined as $G(a,B,C) = \Lambda(a,B,C) \cup
H (a,B,C).$

    \vspace{0.5cm}

 In order to make $G(a,B,C)$ an NW-family, we need to define the $v$-local rank
labels of the $v$-outgoing edges for each vertex $v$ of
$G(a,B,C)$. This can be accomplished by ordering lexicographically
the elements of $\red{A(a,B,C)}$ and ranking them in the ascending
order: $0,1,\ldots,|\red{A(a,B,C)}|-1$. The edge
$(s,\red{\alpha})$ gets as $s$-local rank the same rank as
$\red{\alpha}$. An edge of type $(\red{\alpha}, \ceila{\alpha})$
gets $\red{\alpha}$-local rank $0$, because it is the unique
$\red{\alpha}$-outgoing edge. For $\gamma \in V(H(a,B,C))$ we have
already defined the $\gamma$-local label-ranks. With these local
ranks the two conditions of NW-family are satisfied by $G(a,B,C)$.
It remains to verify that its $st$-paths encode the elements of
$A(a,B,C)$:

\begin{teorema} [Main Theorem] For every phorma $P=(a,B,C)$ the $st$-paths
of $G(a,B,C)$ are in $1-1$ correspondence with the elements of
$A(a,B,C)$. \label{theorem:mainTheorem}
\end{teorema}

\comecaprova  Given an $\alpha \in A(a,B,C)$, let
$\ceila{\alpha}=\gamma^\star$. Define $\pi_\alpha=
(s,\red{\alpha}) \circ (\red{\alpha},\ceila{\alpha}) \circ
\pi_{\se{\alpha}}$. Reciprocally, given an $st$-path $\pi$ in
$G(a,B,C)$, let $\beta$ be the second vertex of $\pi$,
$\gamma^\star$ be its third vertex and $\gamma$ be such that $\pi
= (s,\beta) \circ (\beta,\gamma^\star) \circ \pi_\gamma$. Define
$\alpha_\pi \equiv (\beta,\gamma)$. These definitions imply
$\pi_{\alpha_\pi}=\pi$ and $\alpha_{\pi_\alpha}=\alpha$.
\terminaprova

Given $\red{A(a,B,C)}$ ordered lexicographically and $\beta \in
\red{A(a,B,C)}$ define $||\beta|| = \sum \{ |\beta'| {\rm \ such \
that \ } \beta' < \beta\}$. In Figure \ref{fig:G7575BL} the values
of $||\beta||$ appear as the second number near each vertex
$\beta$. The hash function $h$ for a phorma  assumes a
particularly simple expression:

\begin{proposicao} Given $\alpha \equiv (\red{\alpha},\se{\alpha}) \in A(a,B,C)$, the
perfect hash function $h$ of $G(a,B,C)$ is
$$h(\alpha)=||\red{\alpha}|| + h_{\lceil \alpha \rceil ^a}
(\se{\alpha}).$$
\end{proposicao}
\comecaprova This value of $h(\alpha)$ follows from the general
algorithm for ranking in an abstract NW-family, when specialized
to phormas. \terminaprova

\section{Implementation Aspects}

The need of the boolean function $B$ in a phorma $(a,B,C)$ is just
to enable the enumeration of $\red{A(a,B,C)}$. If the size of this
set is small, then an explicit list of its elements, $\{\beta^1,
\beta^2, \ldots, \beta^u\}$, can be given in place of $B$. If this
is not the case, then a convenient way to input a generic phorma
type boolean function is by means of a tree $T(B)$ with three
types of internal nodes: $\vee$-nodes, $\wedge$-nodes,
$\neg-$nodes. The leaves of the tree correspond to the basic
constituent boolean functions of type $\alpha_i \star \alpha_j$,
where $\star \in \{\le,\ge,<,>,=,\ne\}$. The $\neg-$nodes
(negation operator) must have at most one child. Note that each
subtree rooted at an internal $\diamond$-node $v$ $(\diamond \in
\{\vee, \wedge, \neg\})$  is a boolean tree obtained by taking the
$\diamond$-operation of the boolean tree(s) corresponding to the
children of $v$.  Given an $\alpha$, it is possible to decide its
$B$-satisfiability, by evaluating from the leaves up and arriving
to the root of $T(B)$. See \cite{Booch1991} for more details.

We also admit two ways of inputting $C$: by means of an explicit
list of its elements, $\{\delta^1, \delta^2, \ldots, \delta^z\}$,
if $z=|C|$ is small, or by a phorma type of boolean restrictions
on the coordinates of the $\delta$'s. In this case, $C$ is itself
a boolean expression with clauses of type $(\delta_i \star
\delta_j)$. In the case $C=C^n$, this boolean expression is empty.
We define an NW-family encoding $\bigcup_{n\in
\nat^\star}\{C^n\}$: consider the digraph $L^\infty$, whose vertex
set is the set of points in the plane which have positive integer
coordinates. There are at most two edges from a point $(p,q) \in
V(L^\infty)$, namely a {\em west edge} $((p,q), (p-1,q))$, if
$p\ge 2$, and a {\em southwest edge} $((p,q), (p-1,q-1))$, if
$p,q\ge 2$. The $(p,q)$-local rank-label of the first edge is $0$,
if it exists, and the $(p,q)$-local rank-label of the second edge
is $1$, if both edges exist. In the case that only the second edge
exists, then its $(p,q)$-local rank-label is $0$. Let $C^n_m$ be
the subset of $C^n$ of $n$-compositions which have length $m$.

\begin{proposicao} The paths from $(n,m)$ to $(1,1)$ in $L^\infty$
are in $1-1$ correspondence with the elements of $C^n_m$. Thus
$L^\infty$ is an NW-family encoding the $n$-compositions for all
$n \in \nat$. \label{proposition:NWFamilyOfmnCompositions}
\end{proposicao}

\comecaprova Let $\delta \in C^n_m$ be given. Construct a path
$\pi_\delta$ from $(n,m)$ to $(0,0)$ in $L^\infty$ as follows. Let
$\delta' \leftarrow \delta$ and $\pi' \leftarrow$ the empty path.
Repeat $n$ times: if $\delta'_1>1$, then $\delta'_1 \leftarrow
\delta'_1-1$, extend $\pi'$ with a west edge; if $\delta'_1=1$,
then $\delta'$ becomes $\delta'$ without its first part; extend
$\pi'$ with a southwest edge. After the $n$ iterations of this
loop, $\delta'$ is the composition $1$ of $1$ in $1$ part and
define $\pi_\delta=\pi'$. Reciprocally, given a path $\pi$ from
$(n,m)$ to $(1,1)$ in $L^\infty$, construct a $\delta_\pi \in
C_m^n$ as follows. Let $\delta' \leftarrow 1$ and $\pi' \leftarrow
\pi$. For $i=1,2,\ldots,n$ do: if the $i$-th edge of $\pi$ is a
southwest edge, let $\delta' \leftarrow (1,\delta')$; if the
$i$-th edge of $\pi$ is a west edge, let $\delta'_1 \leftarrow
\delta'_1+1$. Define $\delta_\pi=\delta'$. These definitions imply
$\delta_{\pi_\delta}=\delta$ and that $\pi_{\delta_\pi}=\pi$,
establishing a $1-1$ correspondence between $C^n_m$ and the paths
from $(n,m)$ to $(1,1)$ in $L^\infty$. \terminaprova

By using Proposition \ref{proposition:NWFamilyOfmnCompositions} it
is possible to generate in an efficient way the $\delta$'s
satisfying the boolean expression $C$ via a $C$-restricted
implicit enumeration based on $L^\infty$.

The crucial task to construct (at compiler time) the digraph
$G(a,B,C)$ is to explicitly generate $\red{A(a,B,C)}$. Since
$\alpha$ and $\red{\alpha}$ are order isomorphic, one possibility
to produce $\red{A(a,B,C)}$ is to generate all the $n^n$ members
of $N^N$ and to test each such sequence for reducibility and
$B$-satisfiability \cite{CormenLeisersonRivest1990}. This simple
minded approach is suitable for small dimension $n$. In our
$4$-dimensional phorma $((7,5)^2,B_L,C^4)$ there are only $256$
tests to be made. When $n$ increases this simple minded method
becomes inapplicable. For example, for the $7$-phorma
$(a,B_T^z,C^7)$ there are $7^7=823,543$ tests to be made and a
better approach is needed to generate the $1,134$ elements of
$\red{(A(a,B_T^z,C^7)}$ as well as the $20$ elements of
$\lceil{A(a,B,C^7)}\rceil^a$ (for $a=(15^217^219^3)$). The basic
idea is to implement a {\em $B$-restricted} implicit enumerating
scheme {\em which takes into account only reduced sequences} in
generating the set $\red{(A(a,B,C)}$. This methodology extends
substantially the realm of the phorma applicability.

Given a phorma $(a,B,C)$ and $\delta \in C$. Define
$$\red{A(a,B,C,\delta)}= \{ \alpha \in
\red{A(a,B,C)} \ | \ \o{\alpha}=\o{\red{\alpha}}=\delta\}.$$ As we
know how to generate $\delta \in C$, the generation of
$\red{A(a,B,C)}$ reduces to the generation of each
$\red{A(a,B,C,\delta)}$, because ($\dot \cup$ means disjoint
union)
$$\red{A(a,B,C)} = \dot{\cup}_{\delta \in C} \red{A(a,B,C,\delta)}.$$

The {\em $m$-dimensional grid digraph $J^m$} is the digraph whose
vertices are the points of $I\hspace{-1.5mm}R^m$ with integer
coordinates. There is an edge from $p=p_1 \ldots p_j\ldots p_m$ to
$q=q_1 \ldots q_j \ldots q_m$ if $p_i=q_i$ except for $i = j$,
where $p_j=q_j+1$.

\begin{proposicao}
\label{proposicao:griddigraphJ} An element of
$\red{A(a,B,C,\delta)}$ corresponds to a path from the point
$\delta$ to the origin in $J^{m_\delta}$.
\end{proposicao}
\comecaprova Given $\beta=\beta_1\beta_2\ldots\beta_{m_\delta} \in
\red{A(a,B,C,\delta)}$ we define a path named $\pi_\beta$ in
digraph $J^{m_\delta}$ from $\delta$ to the origin as follows.
Path $\pi_\beta$ starts at $\delta$ and its $i$-th edge is the
edge parallel to the $\beta_i$-th axis. It follows from the
definitions that $\pi_\beta$ finishes at the origin. \terminaprova

From Proposition \ref{proposicao:griddigraphJ} a $B$-restricted
implicit enumeration scheme based on paths in $J^m$, only produces
reduced words. The construction of $\red{A(a,B,C)}$, $\lceil
A(a,B,C\rceil ^a$, and as a consequence, the construction of the
digraph $\Lambda(a,B,C)$ are efficiently performed in this way.

Now we turn our attention to the construction and storage of the
digraph $H(a,B,C)$. Let $\L(r,m) = \{ \gamma \in V(H(a,B,C)) \ | \
\gamma \in (\nat^\star)^M, \ \gamma_m=r\}$ and
$\lceil{A(a,B,C)}\rceil^a_{max} = \{ \gamma^\star \in
\lceil{A(a,B,C)}\rceil^a,\ \gamma^\star {\rm \ maximal} \}$.

\begin{proposicao} $|\ \L(r,m)| \le |\lceil{A(a,B,C)}\rceil^a_{max}|$.
\label{proposition:verticalListBound}
\end{proposicao}
\comecaprova For each element $\gamma \in \L(r,m)$ choose some
$\gamma^\star \in \lceil{A(a,B,C)}\rceil^a_{max}$ such that
$\gamma \preceq \gamma^\star$ . This defines a function $f$ from
$\L(r,m)$ to $\lceil{A(a,B,C)}\rceil^a_{max}$, given by
$f(\gamma)=\gamma^\star$. It is enough to prove that $f$ is
injective. Let $\gamma$ and $\gamma'$ be distinct elements of
$\L(r,m)$. Note that $\sw{\gamma} \neq \sw{\gamma}'$. Suppose that
$\o{\gamma}$ and $\o{\gamma}'$ are such that  $\gamma \preceq
\o{\gamma}$ and $\gamma' \preceq \o{\gamma}'$. Then it follows
that $\o{\gamma} \neq \o{\gamma}'$ because the first $m-1$ entries
of $\o{\gamma}$ form $\sw{\gamma}$ and the first $m-1$ entries of
$\o{\gamma}'$ form $\sw{\gamma}'$. So, $f$ is injective.
\terminaprova

Let $a_\star = \max \{a_i\}$, $n_\star = \max \{ m \ |\ \exists\
\delta \in C \ {\rm \ with\ } m_\delta=m\}$ and $\nu$ the number
of non-empty $\L(r,m)$'s.

\begin{proposicao} $|V(H(a,B,C))| \le 1+ |\lceil{A(a,B,C)}\rceil^a_{max}|(a_\star -(n_\star-1)/2)n_\star$.
\label{proposition:BoundingVH}
\end{proposicao}
\comecaprova Clearly,\ $\nu \le (a_\star -(n_\star-1)/2)n_\star$.
The term $1$ is for the sink $t$. The inequality follows from
Proposition \ref{proposition:verticalListBound}. \terminaprova

The bound given in Proposition \ref{proposition:BoundingVH} is not
tight. In general, the maximum value of $|\ \L(r,m)|$, $\lambda$,
tends to be much smaller than $|\lceil{A(a,B,C)}\rceil^a_{max}|$.
A more informative parameter related to the size of $H(a,B,C)$ is
$\mu$ defined as
$$\mu=|V(H(a,B,C))|/\nu.$$ For phormas arising in the realm of the
applications that we have explored, $\mu$ is rather small. Given a
vertex $\gamma$ of this digraph, $\sw{\gamma}$ and $\w{\gamma}$
are easily obtainable. So the edges of $H(a,B,C)$ do not need to
be stored. Each one of the $\nu$ $\ \L(r,m)$'s is kept as a
lexicographically ordered list indexed by an $(a_\star \times
n_\star)$-array. The $(r,m)$-entry of this array is a pointer to
the list $\L(r,m)$. A binary search can then be used to locate a
specific member of $\L(r,m)$, when computing $h$ and $h^{-1}$.

The amount of work needed to compute $h(\gamma)$ is basically
proportional to $m_\gamma$, the length of $\gamma$. Indeed, from
Proposition \ref{proposition:RankingUnranking} we need only to
find the $m$ elements of the set $PostFall(\pi_\gamma)$ and add
their orders. These orders are stored at the construction of
$H_\gamma$. This makes the time for computing $h(\alpha)$
independent of $a_\star$.

Figure \ref{fig:GParameters} displays basic parameters of various
phormas. The following shortcuts are used: $v_G=|V(G(a,B,C))|$,
$v_H=|V(H(a,B,C))|$, $\alpha^a = |\ceila{A(a,B,C)}|$,
$\alpha^a_{max} = |\ceila{A(a,B,C)}_{\max}|$. The last column of
Figure \ref{fig:GParameters} is $10^4 \times d$, with
$d=|A(a,B,C)|/$\break $(\prod_{i \in N} a_i)$ the {\em density} of
$(a,B,C)$. It is interesting to observe how fast the densities of
the symmetric phormas (the ones with $B=B^{n\ge}_{sym}$) go to
zero as $n$ increases. We present parameters for the phormas
$(9^n,B^{n>}_{sym},C^n)$, $2 \le n \le 9$. The boolean functions
$B^{n}_{sym}$ for these phormas are obtained from $B^{n\ge}_{sym}$
by replacing the inequalities $\ge$ by the strict inequalities
$>$. Thus, only strictly decreasing sequences are permitted. Note
that $A(9^{10},B^{n>}_{sym},C^{10})=\emptyset$.

    \vspace{0.5cm}
    \begin{figure}
        \begin{center}
        {\small
             $\begin{array}{|c|c|c|c|c|c|c|c|c|c|} \hline
             {\rm Phorma} & v_G & v_H &  |\red{A}| & |A| & \alpha^a&
               \alpha^a_{max} & \lambda& \mu & 10^4d \\ \hline
             9^2B_{sym}^{2\ge}/B_{sym}^{2>}C^2 & 20/19 & 17/17 & 2/1 & 45/36 & 2/1 & 1/1 & 1 &  1.0000 & 5556/4444 \\ \hline
             9^3B_{sym}^{3\ge}/B_{sym}^{3>}C^3 & 29/24 & 24/22 & 4/1 & 165/84 & 3/1 & 2/1 & 1 &  1.0000 & 2263/1152 \\ \hline
             9^4B_{sym}^{4\ge}/B_{sym}^{4>}C^4 & 39/27 & 30/25 & 8/1  & 495/126 & 4/1 & 3/1 & 1 &  1.0000  & 754/192\\ \hline
             9^5B_{sym}^{5\ge}/B_{sym}^{5>}C^5 & 52/28 & 35/26 & 16/1 & 1287/126 & 5/1 & 4/1 & 1 &  1.0000  & 218/21\\ \hline
             9^6B_{sym}^{6\ge}/B_{sym}^{6>}C^6 & 72/27 & 39/25 & 32/1 & 3003/84 & 6/1 & 5/1 & 1 &  1.0000  & 57/1.58 \\ \hline
             9^7B_{sym}^{7\ge}/B_{sym}^{7>}C^7 & 107/24 & 42/22 & 64/1 & 6435/36 & 7/1 & 6/1 & 1 &  1.0000  &13/0.0752 \\ \hline
             9^8B_{sym}^{8\ge}/B_{sym}^{8>}C^8 & 173/19 & 44/17 & 128/1 & 12870/9 & 8/1 & 7/1 & 1 &  1.0000 &3/0.0030  \\ \hline
             9^9B_{sym}^{9\ge}/B_{sym}^{9>}C^9 & 302/12 & 45/10 & 256/1 & 24310/1 & 9/1 & 8/1 & 1 &  1.0000  &0.6/10^{-5} \\ \hline
             9^{10}B_{sym}^{10\ge}C^{10} & 557 & 45 & 511 & 43758 & 9 & 8 & 1 &  1.0000 & 0.1 \\\hline\hline

              (7,5)^2B_LC^4 & 32 & 22 & 9 & 190 & 7 & 4 & 2 &  1.0476  & 1551 \\ \hline
              (40,30)^2B_LC^4 & 164 & 154 & 9 & 245670 & 7 & 4 & 2 &  1.0621  & 1706 \\ \hline
              (50,40)^2B_LC^4 & 204 & 194 & 9 & 652910 & 7 & 4 & 2 &  1.0486  & 1632 \\ \hline
              (60,50)^2B_LC^4 & 244 & 234 & 9 & 1420325 & 7 & 4 & 2 &  1.0400  & 1578 \\ \hline
              (99,50)^2B_LC^4 & 400 & 390 & 9 & 5196500 & 7 & 4 & 2 &  1.1404 & 2121 \\\hline
              (100,50)^2B_LC^4 & 404 & 394 & 9 & 5317825 & 7 & 4 & 2 &  1.1420 & 2127 \\\hline\hline

              10^7B_T^zC^7 & 1184 & 49 & 1134 & 237325 & 7 & 6 & 1 & 1.0000 & 237\\ \hline
              15^7B_T^zC^7  & 1219 & 84 & 1134 & 3853200 & 7 & 6 & 1 & 1.0000 & 226 \\ \hline
              20^7B_T^zC^7 & 1254 & 119 & 1134 & 28226800 & 7 & 6 & 1 & 1.0000 & 221\\
              \hline

              25^7B_T^zC^7 & 1289 & 154 & 1134 & 132916875 & 7 & 6 & 1 & 1.0000 & 218 \\ \hline

              30^7B_T^xC^7 & 1324 & 189 & 1134 & 472460925 & 7 & 6 & 1 & 1.0000 & 216\\ \hline
              15^217^219^3B_T^zC^7 & 1262 & 127 & 1134 & 7510130 & 20 & 13 & 3 & 1.1651 & 168\\ \hline
              25^227^229^3B_T^zC^7 & 1332 & 197 & 1134 & 204089675 & 20 & 13 & 3 & 1.1006 & 184\\ \hline
              10^250^212^3B_T^zC^7 & 1201 & 66 & 1134 & 390270 & 17 & 10 & 2 & 1.0645 & 9\\ \hline \hline

              3^{81}B^{Cube}_{It}c^3_{27} & 1013 & 4 & 1008 & 1008 & 1 & 1 & 1 & 1.0000 & 10^{-32}\\ \hline
            \end{array}$}
            \caption{\sf Parameters values for $G(a,B,C)$ of various phormas}
            \label{fig:GParameters}
        \end{center}
    \end{figure}

\section{Conclusion} We have defined a data structure generator
which permits the perfect hash of order restricted
multidimensional arrays $A(a,B,C)$. The restrictions accord a
general type of boolean functions $B$ formed by order restricting
pairs of entries of the array in arbitrary ways. The boolean
function $B$ is used in forming a reduced set $\red{A(a,B,C)}$,
inducing a partition of $A(a,B,C)$. An $\red{\alpha} \in
\red{A(a,B,C)}$ corresponds to a member subset
$\red{A(a,B,C,\o{\red{\alpha}})}$ of this partition. The elements
of $\red{A(a,B,C,\delta})$, $\delta \in C$, are in $1-1$
correspondence with paths from $\delta$ to the origin in the
$m_\delta$-dimensional integer grid digraph $J^{m_\delta}$, and
can be efficiently found in a $B$-restricted implicit enumeration
scheme which produces only reduced sequences. To generate all $c
\in C$, which might be itself a boolean function on the
$\delta$'s, we use the NW-family $R^\infty$ in a $C$-restricted
implicit enumeration search. The whole scheme is summarized by two
facts: (i) an $\alpha \in A(a,B,C)$ induces three pieces of
information, $\red{\alpha}$, $\se{\alpha}$ and  $\lceil\alpha
\rceil ^a$ and is recoverable from the first two, $\alpha \equiv
(\red{\alpha}, \se{\alpha})$; (ii) this decomposition reflects in
the rank formula for a perfect hashing of $A(a,B,C)$:
$h(\alpha)=||\red{\alpha}|| + h_{\lceil \alpha \rceil
^a}(\se{\alpha}).$ This encoding scheme has the power of perfectly
addressing huge and quite intricate arrays $A(a,B,C)$ by means of
the logarithmically smaller NW-family $G(a,B,C)$. This general
type of perfect hash scheme does not seem to have been treated
before in the literature. In particular, its use in database
systems is a possible source of relevant applications and remains
to be investigated.

\vspace{-8mm}

\section{Acknowledgements}
\vspace{-8mm}
The authors thank A. Bondy for bringing reference
\cite{Hoffman1981} to their attention. They also thank three
anonymous referees for helpful comments improving the legibility
of the paper. The financial support of CNPq, (contract no.
30.1103/80) in the case of the second author, is acknowledged.

e-mail addresses: ldl@cin.ufpe.br, sostenes@dmat.ufpe.br,
silvio@dmat.ufpe.br

\end{document}